# Zero-Effort Two-Factor Authentication Using Wi-Fi Radio Wave Transmission and Machine Learning

Ali Abdullah S. AlQahtani
*Computer Systems Technology*
North Carolina A&T State University
Greensboro, North Carolina, USA
alqahtani.aasa@gmail.com

Thamraa Alshayeb
*Physics*
North Carolina A&T State University
Greensboro, North Carolina, USA
alshayeb.t@gmail.com

*Abstract*—The proliferation of sensitive information being stored online highlights the pressing need for secure and efficient user authentication methods. To address this issue, this paper presents a novel zero-effort two-factor authentication (2FA) approach that combines the unique characteristics of a user's environment and Machine Learning (ML) to confirm their identity. Our proposed approach utilizes Wi-Fi radio wave transmission and ML algorithms to analyze beacon frame characteristics and Received Signal Strength Indicator (RSSI) values from Wi-Fi access points to determine the user's location. The aim is to provide a secure and efficient method of authentication without the need for additional hardware or software. A prototype was developed using Raspberry Pi devices and experiments were conducted to demonstrate the effectiveness and practicality of the proposed approach. Results showed that the proposed system can significantly enhance the security of sensitive information in various industries such as finance, healthcare, and retail. This study sheds light on the potential of Wi-Fi radio waves and RSSI values as a means of user authentication and the power of ML to identify patterns in wireless signals for security purposes. The proposed system holds great promise in revolutionizing the field of 2FA and user authentication, offering a new era of secure and seamless access to sensitive information.

## I. INTRODUCTION

User authentication is a crucial aspect of computer security that verifies the identity of an individual attempting to access a system or network. The process of authentication can take many forms, ranging from simple password-based systems to complex biometric methods that utilize physical characteristics such as fingerprints or facial recognition. To enhance security, 2FA combines two different methods of identity verification, such as a password and a one-time code sent to a device. However, traditional 2FA can pose a challenge for users, especially when attempting to access systems on-the-go. To address this issue, 0E2FA has been developed to provide a more convenient, secure, and efficient solution for user authentication.

The proposed 0E2FA system leverages Wi-Fi radio waves, RSSI values, and ML to facilitate the authentication process. The system employs Wi-Fi radio waves, which are electromagnetic radiation commonly used for wireless communication, to determine a user's location by analyzing overlapping access points detected by both of the user's devices. The Service Set Identifier (SSID) and Basic Service Set Identifier (BSSID) of each detected access point are analyzed to identify the user's location. RSSI values, a metric used to measure the strength of a wireless signal, are employed to determine the user's access privileges to a system or network by applying ML techniques to analyze the values. ML is a subfield of Artificial Intelligence that enables systems to learn and improve from data. The proposed 0E2FA system uses ML to identify patterns in a user's Wi-Fi signal and recognize distinct characteristics that are unique to a specific user. This information is then used to authenticate the user automatically, offering a practical, secure, and efficient way to authenticate users.

The 0E2FA application is significant in various industries, including finance, healthcare, and retail. In finance, it allows customers to access their accounts without the need for lengthy passwords or security questions. In healthcare, it ensures that only authorized personnel have access to patient records. In retail, it provides a secure and seamless shopping experience for customers. As security continues to play a crucial role in different sectors, 0E2FA is likely to become more prevalent in the future.

The proposed 0E2FA system aims to enhance user authentication by incorporating unique aspects of the environment through Wi-Fi radio wave transmission and ML. This approach leverages the Wi-Fi radio waves and received signal strength indicator (RSSI) values to determine the user's location and employs ML techniques to analyze the data. As a result, the system can verify the user's identity automatically without the need for additional hardware or software. This results in a more secure and user-friendly authentication process.

This paper presents a novel approach to user authentication through the use of 0E2FA, which incorporates Wi-Fi radio wave transmission and ML. It provides a practical solution for improving the security of sensitive information in various industries. The effectiveness of this approach is demonstrated through a working prototype and experimentation. Furthermore, the research sheds light on the potential of Wi-Fi radio waves and RSSI values for user authentication and the effectiveness of ML in identifying patterns in wireless signals for security purposes. The main contribution of this paper is to present a new and efficient method of user authentication through 0E2FA.

## II. RELATED WORK

The concept of 0E2FA has been extensively examined in various industries as a way to secure sensitive information and systems. 2FA involves using at least two different methods of verifying a user's identity, which improves the security of the system. In this section, we will examine related research on the utilization of 2FA in various contexts.

The widespread use of IoT devices has resulted in a significant increase in the amount of sensitive information being produced. These devices are at risk of spoofing or impersonation attacks, which can be prevented through the implementation of 2FA. Research has shown that 2FA is an effective method for securing IoT systems from these types of attacks, as demonstrated in various studies (e.g. [1]–[6]). One specific example, [7], presents a 2FA protocol for IoT devices that uses physical unclonable functions and the unique characteristics of the device's wireless signal to protect against spoofing and other attacks. This proposed protocol was tested on MICAz motes and found to be effective in securing IoT systems, while also having a lower computational cost and energy consumption than other available methods. The high number of IoT devices in use and the potential sensitivity of the data they produce make them vulnerable to impersonation attacks, and the proposed 2FA protocol provides a straightforward and cost-efficient solution to this problem.

In the healthcare industry, 2FA is becoming increasingly necessary to safeguard patient information, especially with the rise of electronic health records (EHRs). Previous research has focused on creating authentication methods to prevent unauthorized access to patient data [8]–[15]. A recent study [16] proposed a privacy-preserving three-factor authentication key agreement scheme for use in electronic healthcare systems. This scheme aims to provide mutual authentication, session key negotiation, and privacy protection with minimal computational cost. The proposed scheme has been proven to be secure in the Real- or-Random model and has been found to have a lower computational cost in experiments using Raspberry Pi compared to three other related schemes. This scheme is particularly useful during the current COVID-19 pandemic as it is lightweight and suitable for use with resource-constrained devices. Additionally, telehealth systems that allow people to receive medical care remotely using smart devices and 5G networks have become increasingly popular.

A new protocol called Two-Factor Lightweight Privacy-preserving Authentication (2FLIP) has been developed to improve the security of communication in vehicular ad-hoc networks (VANETs) [17]. This protocol combines decentralized certificate authority with a biological-password-based 2FA to provide strong privacy preservation, nonrepudiation, and resistance to denial-of-service attacks. Unlike previous schemes, 2FLIP has significantly lower computational and communication overhead and is able to handle complex transportation situations. Simulation results indicate that the 2FLIP protocol has a very low network delay and packet loss ratio, making it suitable for real-time emergency reporting applications.

According to a study [18], there are potential security risks associated with mobile 2FA applications that use SMS-based authentication. To address this, the researchers propose a solution that involves moving these applications to the cloud, which offers enhanced security and resources. They present an offloading architecture for 2FA applications and a new 2FA scheme that utilizes a virtual smart card. They also outline a method for determining when and how to offload 2FA applications and virtual smart cards based on security, mobile device energy, and cost. The security of the proposed architecture is evaluated and performance results are provided for the 2FA protocol and offloading decision-making process. The goal of this work is to improve the security of mobile 2FA applications by utilizing the cloud and the proposed security measures.

A study [19] introduces a secure three-factor mutual authentication protocol for electronic governance systems operating in multi-server environments. The protocol aims to address the weaknesses of a previous smart card-based authentication method by providing resistance to smart card theft, insider attacks, man-in-the-middle attacks, user impersonation, and session key disclosure. The protocol's security is evaluated through informal analysis, Burrows-Abadi-Needham logic, Real-or-Random modeling, and simulations using the Automated Validation of Internet Security Protocols and Applications tool. It is also compared to a similar protocol ([20]) in terms of security functionality, computation costs, and communication overhead, and is found to be secure and suitable for use in electronic governance systems.

In summary, research has shown that using 2FA is an effective way to enhance security and prevent impersonation attacks across various fields, including IoT devices, healthcare systems, vehicular ad-hoc networks, and mobile applications. However, more research and development is needed to improve the security and effectiveness of these protocols. The next section of this paper presents the proposed system that uses ML to predict a user's location based on beacon frame characteristics and RSSI values.

## III. THE PROPOSED SCHEME

In this section, we present the proposed scheme for a 0E2FA system. The goal of this system is to confirm the identity of a user by utilizing the unique features of their surroundings through Wi-Fi radio wave transmission and ML techniques. We start by providing an overview of the system's architecture, including the specific roles of each component. Then, we describe the procedure for verifying a user's identity through the access policy. Finally, we delve into the specifics of the authentication process of the proposed system.

## A. System Architecture

The system architecture of the proposed 0E2FA system is described in this section. We outline the various components of the system and their respective roles and functions. Understanding the system architecture is crucial for comprehending the operation and identity verification process of the proposed 0E2FA system.

*1) Radio wave transmission:* The proposed 0E2FA system uses radio waves transmitted by Wi-Fi access points (APs) to authenticate a user based on the distinctive characteristics of their surroundings. The system requires the presence of at least one Wi-Fi AP in the user's environment. The Wi-Fi AP transmits beacon frames that contain unique features such as the SSID, BSSID, and frequency. These beacon frame characteristics, specific to the user's environment, are used to verify the user's identity.

*2) User's two Devices:* The proposed system mandates that the user possesses two Wi-Fi enabled devices, such as a smartphone and a laptop. These devices must be capable of receiving, collecting data from, and transmitting it back to a server. The system uses the unique characteristics of the user's environment, including beacon frame characteristics and received signal strength indicator (RSSI) values, to confirm the user's identity. A smartphone can serve as a mobile device, while a laptop or personal computer can act as a login device. Both devices are essential for the proper functioning and identity verification of the proposed system.

*3) Software requirements:* The proposed 2FA system necessitates software installed on the user's devices that can transmit and receive beacon frames, as well as measure RSSI values. The software is vital for the system to authenticate the user's identity based on the unique characteristics of their environment, including the beacon frame characteristics and RSSI values collected by the user's devices. The software should also be capable of transmitting the collected data back to a server for analysis and identification of the user. The proposed system requires the user to log in through the mobile application during the registration phase in order to activate and create their account. Once the initial login is complete, the application operates in the background and waits for instructions from the authentication entity. Every time the user wants to access the system, the application automatically scans the user's surroundings and sends the collected data to the authentication entity for verification. The user does not need to take any additional steps through the mobile device application during this process.

## B. Server

The proposed system necessitates a central server that acts as the medium for authentication. This server serves as the main point of contact between the user's devices and the entity that requires protection. The server is composed of various components that aid in facilitating the authentication process:

1) A database is utilized to store the beacon frame characteristics and RSSI values that are collected from the user's devices. Additionally, the database also stores the user's login credentials and any other necessary information for the authentication process. This allows the system to verify the user's identity by utilizing the unique characteristics of their surrounding environment.
2) An authentication module that uses ML to analyze the beacon frame characteristics and RSSI values collected from the user's devices to verify the user's identity. The server is responsible for executing this process and making the final decision on whether to grant the user access to the protected entity.

## C. Authentication Process

In this section, we will delve into the specifics of the proposed 0E2FA system's authentication process. The process begins with the user attempting to access the protected entity, which initiates the automatic scanning of the user's surroundings by the mobile device application. The collected beacon frame characteristics and RSSI values are transmitted to the central server for evaluation. The server then compares the data against the information stored in the database, including the user's login credentials and the unique characteristics of their surrounding environment. The authentication module utilizes ML techniques to analyze the data and make a decision on whether to grant access to the protected entity. If the authentication is successful, the user is granted access. If not, the user must repeat the process or use alternative authentication methods. The steps involved in the authentication process are shown in Figure 1.

## IV. EXPERIMENT

The goal of this experiment is to assess the performance of the new 0E2FA system utilizing radio waves and ML to confirm a user's identity. Through the use of ML, this experiment aims to predict a user's location by analyzing beacon frame characteristics (i.e., SSID and frequency) as well as RSSI values. In our study, we utilized various tools to gather data and conduct experiments. Firstly, we utilized all publicly accessible Wi-Fi APs as a source of radio waves without needing to connect to them. Secondly, to simulate user devices, we employed two Raspberry Pi devices as mobile and login devices. Lastly, our server was a desktop computer running Ubuntu 16.04 LTS (64-bit) as the operating system, which hosted the ML module and utilized PhPmyadmin for managing the database.

## A. Data Collection Phase

In the data collection phase of the proposed system, beacon frame characteristics and RSSI values are collected from Wi-Fi APs using two Raspberry Pi devices. The aim of this phase is to acquire sufficient data to train the ML module to accurately identify the user's device location based on these characteristics and values. A threshold distance of 7 feet was established as the maximum distance between the user's devices that is acceptable for the experiment. Two datasets were gathered: one with data collected while the two Raspberry Pis were within 7 feet of each other, and of 7.5 feet or less apart in the *"authentic"* dataset and at distances of 7.5 feet or more apart in the *"unauthorized"* dataset. Each dataset includes six columns: "RPi," "SSID," "Frequency (Hz)," "RSSI (dBm)," "Location," and "Label." The resulting dataset was balanced with 2442 samples in the "authentic" dataset and 2383 samples in the "unauthorized" dataset. The dataset was prepared for the implementation phase of the experiment and it is available on IEEEDataPort [21].

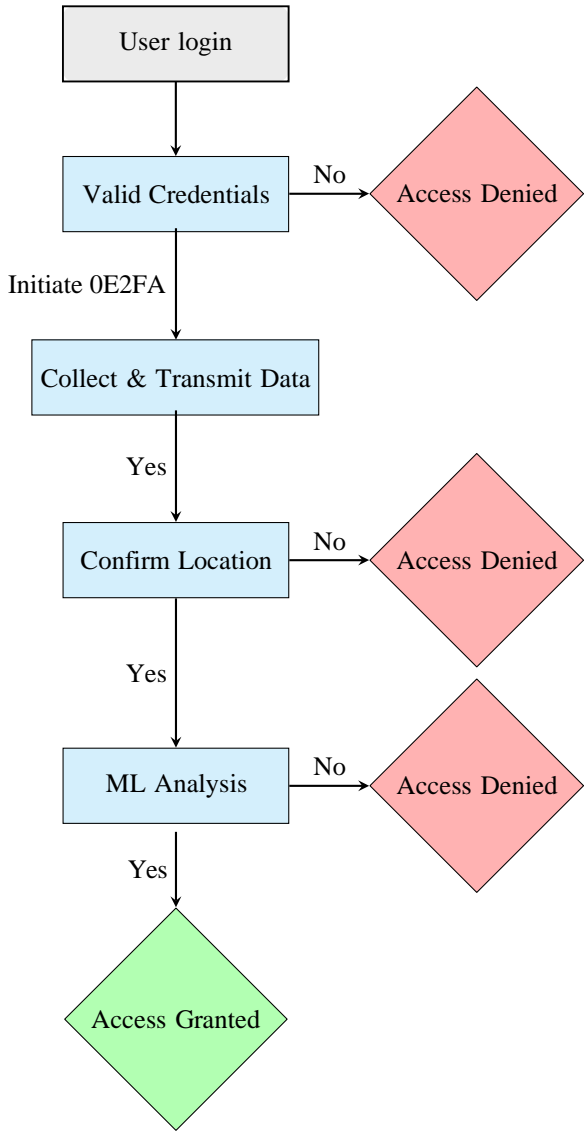

Fig. 1. Authentication Process

*B. Implementation Phase*

In our proposed system, we aimed to utilize ML in order to predict a user's location through analyzing beacon frame characteristics (i.e., SSID and frequency) as well as RSSI values collected from the user's devices. To accomplish this, we employed supervised learning, a type of ML that requires a model to be trained using labeled data. The data used in our experiment was labeled, with samples that fell within our threshold (i.e., 7 feet) being labeled as *"authentic"* and those that exceeded our threshold being labeled as *"unauthorized."* We split the labeled data into a training set (80%) and a testing set (20%) in order to train the models and evaluate the proposed system. Two ML models, Decision Tree (DT) and Random Forest (RF), were used for evaluation.

*C. Evaluation Phase*

After evaluating each of the ML models, we analyzed the results using a confusion matrix. The confusion matrices for the DT and RF models are illustrated in Figures 2 and 3, respectively.

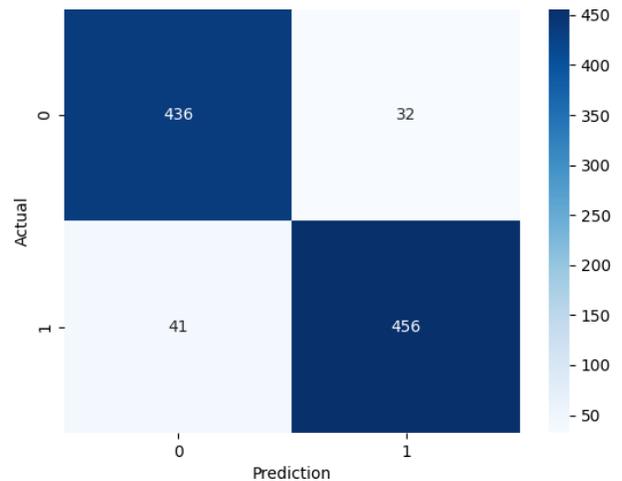

Fig. 2. Decision Tree Confusion Matrix

another with data collected while the distance between the two Raspberry Pis was over 7.5 feet. For the first dataset, the Raspberry Pis were placed 7 feet apart and moved closer and farther while keeping within the threshold distance. The data collection process was repeated at different locations to capture variations in beacon frame characteristics and RSSI values in different environments. For the second dataset, the Raspberry Pis were positioned at a distance of 7.5 feet apart to determine the "gray area" between the acceptable threshold distance and the distance at which access should be denied. The process of moving the devices closer and farther was repeated while keeping the closest distance at 7.5 feet, and the process was repeated in different locations. A total of 4,825 samples of data were collected from two Raspberry Pis capturing the SSID and RSSI values of 10 different WiFi APs at different locations and times. The Raspberry Pis were positioned at distances

Using the information obtained from the confusion matrix, we determined the performance of our models by calculating various evaluation metrics which are accuracy, sensitivity, specificity, precision, and F1 Score. These metrics allow us to evaluate the model's ability to correctly predict the class of a given sample. We used equations (1), (2), (3), (4) and (5) to calculate these metrics, and the results were summarized in Table I.

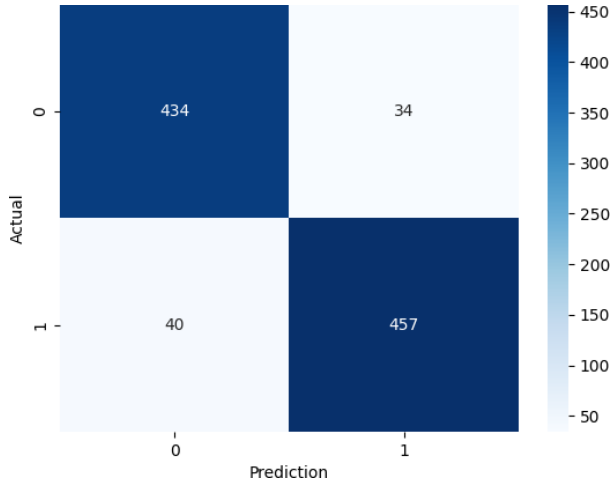

Fig. 3. Random Forest Confusion Matrix

$$Accuracy = \frac{TP + TN}{N} \times 100 \quad (1)$$

$$Sensitivity = \frac{TP}{TP + FN} \times 100 \quad (2)$$

$$Specificity = \frac{TN}{TN + FP} \times 100 \quad (3)$$

$$Precision = \frac{TP}{TP + FP} \times 100 \quad (4)$$

$$F1Score = 2 \times \frac{Precision \times Sensitivity}{Precision + Sensitivity} \quad (5)$$

TABLE I
EVALUATION RESULTS

| Model | Accuracy | Sensitivity | Specificity | Precision | F1 Score |
|---|---|---|---|---|---|
| DT | 0.924 | 0.918 | 0.932 | 0.934 | 0.926 |
| RF | 0.922 | 0.92 | 0.932 | 0.935 | 0.927 |

Table I presents the evaluation results for two different models: DT) and RF. Both models have high overall accuracy, with the DT model having an accuracy of 0.924 and the RF model having an accuracy of 0.922. In terms of sensitivity, the DT model has a slightly higher score at 0.918 compared to the RF model's 0.92. Specificity, which measures the ability of a model to correctly identify negative cases, is similar for both models at 0.932 for the DT model and 0.932 for the RF model. Both models also have high precision scores, with the DT model scoring 0.934 and the RF model scoring 0.935. The F1 score, which is a measure of a model's balance between precision and recall, is also similar for both models at 0.926 for the DT model and 0.927 for the RF model. Overall, both models show strong performance in terms of accuracy, sensitivity, specificity, precision, and F1 score.

The experiments demonstrate that the proposed 2FA system utilizing radio waves and ML is a viable solution for confirming a user's identity. The high level of accuracy achieved in the experiments indicates that the system is highly effective in identifying unauthorized users, and the satisfactory results demonstrate that this system is a reliable solution for securing user access.

## V. CONCLUSION

The need for secure and efficient user authentication methods continues to grow as the amount of sensitive information stored online increases. This paper presents a novel approach for 0E2FA that promises to revolutionize the way we think about user authentication by leveraging the unique characteristics of a user's environment through Wi-Fi radio wave transmission and ML. This approach aims to provide a secure and efficient method of user authentication that does not require additional hardware or software, by utilizing the beacon frame characteristics and RSSI values from Wi-Fi APs to determine a user's location and subsequently, applying ML techniques to verify the user's identity. The proposed approach was evaluated through experiments on a working prototype using Raspberry Pi devices, and the results demonstrated its effectiveness and practicality in improving the security of sensitive information in various industries such as finance, healthcare and retail. Moreover, this research provides insights into the potential of Wi-Fi radio waves and RSSI values as a means of user authentication and the power of ML to identify patterns in wireless signals for security purposes. 0E2FA is likely to become more prevalent in the future as it plays a crucial aspect of security in different sectors. Furthermore, this study can be extended to consider more features or use more sophisticated models that might improve the performance of the proposed system. Future work may also include testing the proposed system in real-world scenarios and evaluating the performance in different environments. The proposed system is a game-changer for the field of user authentication, paving the way for a new era of secure and seamless access to sensitive information.